\documentclass[]{aa}
\usepackage{times,graphicx,amssymb}

\begin{document}
 %
 %
  
\title{Young Stars and Outflows in the globule IC\,1396\,W}

\author{Dirk Froebrich \and
        Alexander Scholz\thanks{Visiting Astronomer, 
German-Spanish Astronomical Centre, Calar Alto, operated by the 
Max-Planck-Institute for Astronomy, Heidelberg, jointly with the Spanish
National Commission for Astronomy.}}

\offprints{A. Scholz, scholz@tls-tautenburg.de}

\institute{Th\"uringer Landessternwarte Tautenburg, Sternwarte 5, 
           D-07778 Tautenburg, Germany}
	
\date{Received sooner / Accepted later}
\authorrunning{D. Froebrich and A. Scholz}
\titlerunning{Young Stars and Outflows in the globule IC\,1396\,W}

\abstract{We have observed the \object{IC\,1396\,W} globule in a narrow band
filter centred on the 1\,--\,0\,S(1) line of molecular hydrogen and in the J,
H, K' broad-band filters. Three molecular hydrogen outflows could be identified
by means of H$_2$ emission. The projected axes of the flows are parallel to
each other. By means of the NIR images and IRAS/ISOPHOT data we could identify
the driving sources of all outflows, the possible Class\,0 source
\object{IRAS21246+5743} and two red objects (Class\,1/2). NIR photometry
reveals an embedded cluster of young stars in the globule, coinciding with FIR
emission. Other young stars in the field are more or less clustered in several
small groups, an indication that star formation takes place at different
positions at the same time in such small globules.

} 

\maketitle


\section{Introduction}

The H\,{\small II} region \object{IC\,1396} is situated in the
\object{Cep\,OB2} region at a distance of about 750\,pc (Matthews \cite{m79}).
It is excited by the O6.5V star \object{HD\,206267} (Walborn \& Panek
\cite{wp84}). Extensive studies of the structure and the content of embedded
young stellar objects in this region were performed by Weikard et al.
(\cite{wwcws96}) and Schwarz et al. (\cite{sgw91}). Osterbrock (\cite{o57}) has
identified several dark globules in this region, while Sugitani et al.
(\cite{sfo91}) proved the association of some of such globules with IRAS
sources. This, and the observations of the \object{IC\,1396\,N} globule (e.g.
Codella et al. (\cite{cbnst01}) and Nisini et al. (\cite{nmvgldc_etal01})) give
strong evidence for the assumption that these globules are sites of ongoing
star formation. 

The IC\,1396\,W globule lies about 1.75$^\circ$ WNW of the star HD\,206267. In
the centre of the small (about 6') dark cloud, the IRAS source IRAS21246+5743
can be found. This source is not detected at 12\,$\mu$m. Together with the very
red IRAS colours and the extended appearance in 100\,$\mu$m IRAS image, this
suggests the existence of a young, deeply embedded source. Further
investigations of this object with the photometer ISOPHOT confirmed this
hypothesis (Froebrich et al. \cite{fshe03}). They find strong evidence for
IRAS21246+5743 being a deeply embedded Class\,0 source of 16\,L$_\odot$ that
will reach about one solar mass on the main sequence. The ISOPHOT maps at 160
and 200\,$\mu$m of Froebrich et al. (\cite{fshe03}) show two further cold
objects (2.5\arcmin\, SW and NE, respectively) in the vicinity of the central
source. This might be an indication of other new forming stars or cold dust in
the IC\,1396\,W globule. Such embedded objects or clusters can be revealed by
NIR imaging in the J, H, and K bands. The resulting (J-H, H-K) colour-colour
diagrams give then additional information about the extinction and the youngest
members of such clusters.

Observations of the IC\,1396\,W cloud in low-excitation rotational CO lines
(see e.g. Wouterloot \& Brand (\cite{wb89}) and Schwartz et al. (\cite{sgw91}))
indicate line wings on the profiles. This gives evidence for the presence of a
mass outflow, connected to the central IRAS source. But no CO map, flow mass,
or energetics were presented so far. Hodapp (\cite{h94}) imaged the region in
the NIR K' filter and describes a parabola-shaped nebula at the western end of
the outflow, an indication of shock excited bow shaped H$_2$ emission as found
in many other protostellar outflows. NIR imaging in continuum and a narrow band
filter centred at the 1\,--\,0\,S(1) line of molecular hydrogen at
2.122\,$\mu$m can provide a verification of this assumption. Such observations
are a powerful tool to investigate the intensity and morphology of shocked
H$_2$ emission. 

In Sec.\,\ref{data} we present our data and describe the data reduction
process. A special emphasis is placed on the calibration of the J, H, and K'
magnitudes. Our results are shown in Sec.\,\ref{results}. We describe the
discovered outflows in detail and characterise the young stellar population of
the IC\,1396\,W globule by means of a (H-K, J-H) colour-colour diagram.


\section{Observations and Data Reduction}

\label{data}

The NIR survey of IC\,1396\,W was carried out with MAGIC (Herbst et al.
\cite{hbbhmmw93}) mounted at the 2.2-m telescope of the German-Spanish
Astronomical Centre on Calar Alto. The camera possesses a $256\times256$
NICMOS3 array with a pixel scale of 1.6\arcsec/pix in widefield mode. Using a
$3\times3$ dithering pattern, we observed a $13.5\arcmin\times13.5$\arcmin\,
field centred on IRAS21246+5743. During a run from 16th to 19th of September
2002, we obtained images  in the broadband filters J, H, K', and the narrow
band filter NB2122 with a central  wavelength of 2.122\,$\mu$m. The total
exposure times were 192\,s in J and H, 208\,s in K', and 3480\,s in NB2122. In
a photometric night, the 2MASS object \object{2MASSW J0036159+182110}
(hereafter 2M0036+18; Kirkpatrick  et al. \cite{k00}) was observed in the
broadband filters J, H, and Kshort as photometric standard star.

Both, object and standard star frames were flatfielded using skyflats. The
DIMSUM package in IRAF\footnote{IRAF is distributed by National Optical
Astronomy Observatories, which is operated by the Association of Universities
for Research in Astronomy, Inc., under contract with the National Science
Foundation.} was used for sky subtraction and co-addition of the single images.
Co-centering of the single frames was done by using all detectable stars in the
field to obtain a high relative astrometric accuracy (about 1\arcsec).

In order to generate a catalogue of all objects in the field (including the
emission knots), we ran the SExtractor (Bertin \& Arnouts \cite{ba96}) on the
NB2122 mosaic. The relative offsets between the NB2122 and the three broadband
images were determined by measuring the relative positions of  some bright
stars. Applying these offsets to the NB2122 pixel coordinates, we obtained 
object catalogues for the J, H, and K' mosaics. We used the PHOT routine within
the IRAF DAOPHOT package (Stetson \cite{s87}) to perform aperture  photometry
for all registered objects. The aperture radius of 4 pixels safely includes
more than 99\% of the object's flux. The instrumental magnitudes of the
calibration object 2M0036+18 were measured with the same parameters.

With the 2MASS magnitudes of 2M0036+18, we calculated accurate zeropoints for
the J, H, and Kshort band. The K' filter used for the IC\,1396\,W survey has a
response function very similar to the Kshort filter: Essentially, both filters
cover the K-band {\it without} the long wavelength end above 2300\,nm. Compared
with the K-band, this excludes a region with strong atmospheric absorption and,
thus, improves the signal-to-noise ratio. The application of the derived
zeropoints shifts the instrumental magnitudes of the IC\,1396\,W field into the
photometric system of 2MASS. Since the standard star was observed nearly at the
same airmass as the IC\,1396\,W field, this operation also corrects for the
atmospheric extinction. The error caused by the  airmass difference between
standard and science observations is below 1\%.

To distinguish between field stars in the back- or foreground and a possible
young population, we need to compare the derived magnitudes with the main
sequence, e.g. with the infrared colours  given by Bessell \& Brett
(\cite{bb88}, hereafter: BB). The referred empirical main sequence is well
adapted  for our purposes because it uses a specified homogenized JHK system.
To calculate the zeropoint  difference of this system with respect to the 2MASS
system, we integrated the filter curves of  both systems and scaled our
magnitudes to the transmission of the BB bands. 

However, one has to be careful comparing our colours with the main sequence of
BB. The described transformation from 2MASS to BB magnitudes is only valid for
objects with flat spectral energy distribution and, thus, colour zero. Since we
do not a priori know the BB colours for our objects, it is not possible to
perform a colour correction. As noted above, the long wavelength  cutoff of the
Kshort (and the K') filter is $\approx 100$\,nm below the end of the BB K-band.
Hence, compared with the BB system, we will loose K-flux proportional to the
colour of the target. As a consequence, our (H-K) colours will be too low. A
similar effect must be expected for the J-band: In this case, the BB band
excludes flux for $\lambda>1360$\,nm, whereas the cutoff of the 2MASS J-filter
(and the MAGIC J-filter as well) lies at 1420\,nm. Hence, we measure too much
flux  in the J-band for red objects, leading to increased (J-H) colours.
Recapitulating, all red objects will move down to right in the (H-K, J-H)
colour-colour diagram. The interpretation of these colours (see Sect.
\ref{photometry}) must account for this effect.

The K' image was multiplied by the ratio of the NB2122/K' band-passes and then
subtracted from the NB2122 image. In the difference image stars possess zero
flux and the H$_2$ emission line objects can be identified easily. Flux
calibration and conversion to W\,m$^{-2}$ of this difference image was done
using the zero-point of 673\,Jy for the K-band as given by Wamsteker
(\cite{w81}).


\section{Results}

\label{results}

\subsection{JHK Photometry - Stellar population of the globule} 

\label{photometry}

\begin{figure}[t]
\resizebox{8.6cm}{!}{\includegraphics[angle=-90,bb=50 60 560 765]{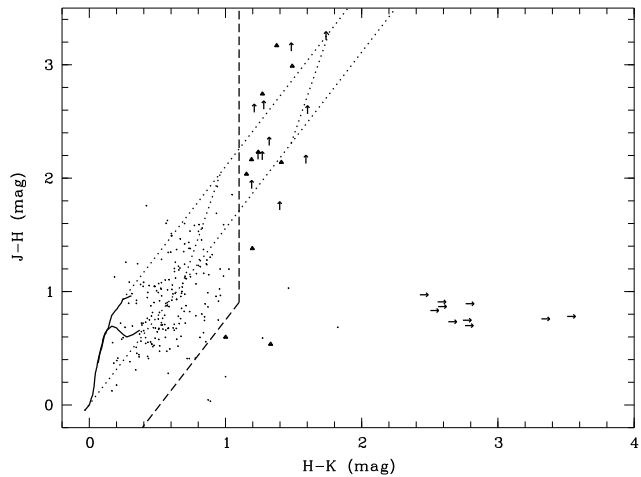}}
\caption{Colour-colour diagram of the observed field. The main sequence and
reddening vectors are indicated. For further explanations see text.}
\label{fhd_ic1396w}
\end{figure}

Our colour-colour diagram for the central part of the observed field is shown 
in Fig.\,\ref{fhd_ic1396w}. Solid lines represent the main sequence for dwarfs
and giants given by BB. The dotted lines indicate the reddening vector
calculated with the extinction model of Mathis (\cite{m90}) and confine the
area where developed stars should reside. Two dotted lines parallel to the main
sequence show the main sequence position for $A_V$\,=\,10 and
$A_V$\,=\,20\,mag. From this diagram, we estimate a mean cloud extinction of
$A_V\approx10$\,mag. As discussed above (see Sect.\,\ref{data}), our photometry
gives too high J-H and too low H-K colours compared with the main sequence as a
consequence of non-homogenuous colour systems. Thus, the objects with very high
J-H can be expected to lie in fact below the extinction path.

Clearly, there is a population of young red objects right of the main sequence.
We selected all objects on the right side of the dashed line as probable young
stellar objects. H$_2$ emission knots and obvious spurious detections were
rejected. The remaining 10 objects are marked as triangles. The reddest objects
will not be detected in J or even in H. We show the position of these
candidates with arrows: Upper arrows represent lower limits in J-H for objects
without detection in J, right arrows are lower limits in H-K for objects
without detection in J and H, but in K' {\it and} NB2122. In these cases, the
J-H values are randomly. We summarise coordinates and photometry of the young
stellar population in IC\,1396\,W in Table\,\ref{redstars}.

\begin{table}[t]
\caption{\label{redstars} Positions and NIR magnitudes and colours of the most
red objects in our observed field. Selection criteria are described in the 
text. Positional errors are in the order of 1\arcsec.}
\begin{center}
\renewcommand{\baselinestretch}{1.0}
\renewcommand{\tabcolsep}{4pt}
{\scriptsize
\begin{tabular}{rccccc}
Object & $\alpha$(J2000) & $\delta$(J2000) & K (mag) & J-H (mag) & H-K (mag) \\
\noalign{\smallskip}
\hline
\noalign{\smallskip}
 1~~~~~&   21:25:41.2 &  57:54:37  & 14.64 $\pm$ 0.08 & 0.60 $\pm$ 0.13 & 1.00 $\pm$ 0.13   \\  
 2~~~~~&   21:25:41.5 &  57:57:18  & 10.75 $\pm$ 0.01 & 3.17 $\pm$ 0.04 & 1.38 $\pm$ 0.01   \\  
 3~~~~~&   21:25:42.9 &  57:55:27  & 14.47 $\pm$ 0.07 & 1.38 $\pm$ 0.20 & 1.20 $\pm$ 0.12   \\  
 4~~~~~&   21:25:47.1 &  57:54:17  & 11.31 $\pm$ 0.01 & 2.74 $\pm$ 0.04 & 1.27 $\pm$ 0.01   \\  
 5~~~~~&   21:25:49.0 &  57:54:42  & 11.62 $\pm$ 0.01 & 2.99 $\pm$ 0.07 & 1.49 $\pm$ 0.01   \\  
 6~~~~~&   21:25:52.8 &  57:55:47  & 13.19 $\pm$ 0.02 & 2.16 $\pm$ 0.12 & 1.19 $\pm$ 0.04   \\  
 7~~~~~&   21:25:54.4 &  57:56:01  & 13.61 $\pm$ 0.03 & 2.23 $\pm$ 0.17 & 1.24 $\pm$ 0.05   \\  
 8~~~~~&   21:26:05.7 &  57:56:44  & 13.04 $\pm$ 0.02 & 2.04 $\pm$ 0.08 & 1.15 $\pm$ 0.03   \\  
 9~~~~~&   21:26:16.6 &  57:55:42  & 15.12 $\pm$ 0.11 & 0.54 $\pm$ 0.26 & 1.33 $\pm$ 0.22   \\  
10~~~~~&   21:26:32.8 &  57:57:45  & 12.42 $\pm$ 0.01 & 2.14 $\pm$ 0.08 & 1.41 $\pm$ 0.02   \\  
11~~~~~&   21:25:45.3 &  57:56:27  & 14.80 $\pm$ 0.08 & $>$2.2~~~~~~~~~~ & 1.24 $\pm$ 0.15 \\  
12~~~~~&   21:25:46.2 &  57:55:45  & 14.77 $\pm$ 0.08 & $>$2.2~~~~~~~~~~ & 1.27 $\pm$ 0.15 \\  
13~~~~~&   21:25:51.6 &  57:56:04  & 15.10 $\pm$ 0.11 & $>$1.9~~~~~~~~~~ & 1.19 $\pm$ 0.20 \\  
14~~~~~&   21:25:52.0 &  57:56:10  & 14.59 $\pm$ 0.07 & $>$2.3~~~~~~~~~~ & 1.32 $\pm$ 0.13 \\  
15~~~~~&   21:25:52.1 &  57:54:58  & 13.24 $\pm$ 0.02 & $>$3.2~~~~~~~~~~ & 1.74 $\pm$ 0.05 \\  
16~~~~~&   21:25:52.6 &  57:56:07  & 14.48 $\pm$ 0.06 & $>$2.1~~~~~~~~~~ & 1.59 $\pm$ 0.14 \\  
17~~~~~&   21:25:59.4 &  57:55:19  & 14.31 $\pm$ 0.05 & $>$2.6~~~~~~~~~~ & 1.28 $\pm$ 0.09 \\  
18~~~~~&   21:26:17.3 &  57:56:30  & 15.08 $\pm$ 0.11 & $>$1.7~~~~~~~~~~ & 1.40 $\pm$ 0.22 \\  
19~~~~~&   21:26:33.7 &  57:55:26  & 13.60 $\pm$ 0.04 & $>$3.1~~~~~~~~~~ & 1.48 $\pm$ 0.07 \\  
20~~~~~&   21:26:40.6 &  57:55:46  & 14.03 $\pm$ 0.06 & $>$2.6~~~~~~~~~~ & 1.60 $\pm$ 0.12 \\  
21~~~~~&   21:26:41.9 &  57:59:30  & 14.41 $\pm$ 0.10 & $>$2.6~~~~~~~~~~ & 1.21 $\pm$ 0.17 \\  
22~~~~~&   21:25:40.4 &  57:54:29  & 15.44 $\pm$ 0.16 &         & $>$2.8~~~~~~~~~~ \\  
23~~~~~&   21:25:44.3 &  57:55:04  & 15.70 $\pm$ 0.19 &         & $>$2.5~~~~~~~~~~ \\  
24~~~~~&   21:25:45.1 &  57:54:22  & 15.64 $\pm$ 0.17 &         & $>$2.6~~~~~~~~~~ \\  
25~~~~~&   21:25:48.7 &  57:53:52  & 15.77 $\pm$ 0.20 &         & $>$2.4~~~~~~~~~~ \\  
26~~~~~&   21:25:49.5 &  57:55:56  & 14.88 $\pm$ 0.10 &         & $>$3.3~~~~~~~~~~ \\  
27~~~~~&   21:26:04.5 &  57:54:47  & 15.64 $\pm$ 0.20 &         & $>$2.6~~~~~~~~~~ \\  
28~~~~~&   21:26:13.1 &  57:56:01  & 15.44 $\pm$ 0.15 &         & $>$2.8~~~~~~~~~~ \\  
29~~~~~&   21:26:23.1 &  57:56:11  & 15.46 $\pm$ 0.15 &         & $>$2.7~~~~~~~~~~ \\  
30~~~~~&   21:26:27.4 &  57:56:19  & 15.56 $\pm$ 0.19 &         & $>$2.6~~~~~~~~~~ \\  
31~~~~~&   21:26:33.9 &  57:55:41  & 14.69 $\pm$ 0.10 &         & $>$3.5~~~~~~~~~~ \\  
\end{tabular}}
\end{center}
\end{table}

\subsection{Outflow morphology and driving sources} 

\label{morphology}

\begin{figure}[t]
\resizebox{8.6cm}{!}{\includegraphics[angle=-90,bb=65 30 530 565]{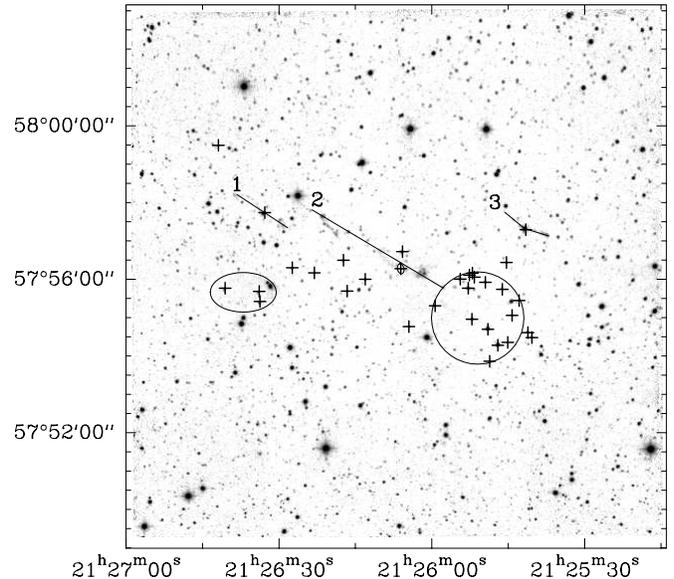}}
\caption{Imaged field in the H$_2$ 1\,--\,0\,S(1) line + continuum. The
detected outflows are marked by a solid line and labeled according to the
numbers used in the text. Possible driving sources, including IRAS21246+5743,
and other red stars (compare Table\,\ref{redstars}) are marked by a cross.
Additionally, two regions containing red sources and discussed in the text are
marked by a circle/ellipse.}
\label{field_ic1396w}
\end{figure}

\begin{table}[t]
\caption{\label{positions}Positions and fluxes of the new discovered
H$_2$-knots. The positional errors are in the order of 1\arcsec. Photometric
errors are about 5 percent for the brighter knots and raise to about 20 percent
for the faint objects. Due to overlap effects the totals of the fluxes for the
whole flows are not identical with the numbers given in
Sect.\,\ref{energy}.}
\begin{center}
\renewcommand{\baselinestretch}{1.0}
\renewcommand{\tabcolsep}{4pt}
{\scriptsize
\begin{tabular}{lccrr}
Object & $\alpha$(J2000) & $\delta$(J2000) & Flux$^*$ & Surface$^{**}$ \\
& & & & Brightness \\ 
\noalign{\smallskip}
\hline
\noalign{\smallskip}
1-a & 21:26:37.3 & +57:58:11 &  8.2 & 1.15 \\
1-b & 21:26:33.8 & +57:57:56 & 12.4 & 1.89 \\
1-c & 21:26:32.1 & +57:57:55 &  4.2 & 0.78 \\
1-d & 21:26:32.8 & +57:57:35 & 10.2 & 1.15 \\
1-e & 21:26:29.6 & +57:57:35 & 24.8 & 5.90 \\
1-f & 21:26:29.9 & +57:57:25 & 13.8 & 2.58 \\
1-g & 21:26:28.9 & +57:57:26 & 24.3 & 4.38 \\
1-h & 21:26:26.7 & +57:57:27 &  6.6 & 1.97 \\
2-a & 21:26:23.5 & +57:57:50 & 12.3 & 3.61 \\
2-b & 21:26:21.2 & +57:57:40 & 65.0 &13.20 \\
2-c & 21:26:20.2 & +57:57:25 & 11.3 & 3.48 \\
2-d & 21:26:18.6 & +57:57:12 & 15.6 & 4.88 \\
2-e & 21:26:16.7 & +57:57:14 &  1.9 & 0.74 \\
2-f & 21:26:09.9 & +57:56:46 & 22.4 & 2.05 \\
2-g & 21:26:06.7 & +57:56:22 & 13.1 & 1.93 \\
2-h & 21:26:04.9 & +57:56:29 & 13.7 & 2.66 \\
2-i & 21:26:05.0 & +57:56:17 & 21.2 & 3.20 \\
2-j & 21:26:01.8 & +57:56:11 &177.3 &16.19 \\
2-k & 21:25:59.1 & +57:56:20 &  6.9 & 1.15 \\
2-l & 21:25:59.5 & +57:55:55 &  6.0 & 1.02 \\
2-m & 21:25:57.7 & +57:55:49 & 11.6 & 2.87 \\
2-n & 21:25:55.4 & +57:55:48 & 11.7 & 3.36 \\
3-a & 21:25:45.5 & +57:57:44 &  3.6 & 1.02 \\
3-b & 21:25:43.8 & +57:57:32 &  6.0 & 1.89 \\
3-c & 21:25:42.5 & +57:57:23 & 14.0 & 3.40 \\
3-d & 21:25:39.6 & +57:57:15 &  2.8 & 1.02 \\
3-e & 21:25:38.0 & +57:57:10 & 21.0 & 5.25 \\
3-f & 21:25:37.2 & +57:57:08 & 29.6 & 9.10 \\
4   & 21:26:19.5 & +57:56:59 &  7.8 & 1.23 \\
5   & 21:26:07.9 & +57:56:05 & 10.1 & 2.79 \\
6   & 21:25:57.6 & +57:55:22 & 16.6 & 1.23 \\
\end{tabular}}
\end{center}
\begin{list}{}{}
\item[$^{\mathrm{*}}$]Fluxes are in $10^{-18}$\,W\,m$^{-2}$.
\item[$^{\mathrm{**}}$]Surface brightness of H$_2$ 1--0\,S(1) averaged over
1.5\,\arcsec\, diameter are in $10^{-19}$\,W\,m$^{-2}$\,arcsec$^{-2}$. 
\end{list}
\end{table}

\begin{figure*}[t]
\resizebox{17.5cm}{!}{\includegraphics[angle=-90,bb=225 70 530 755]{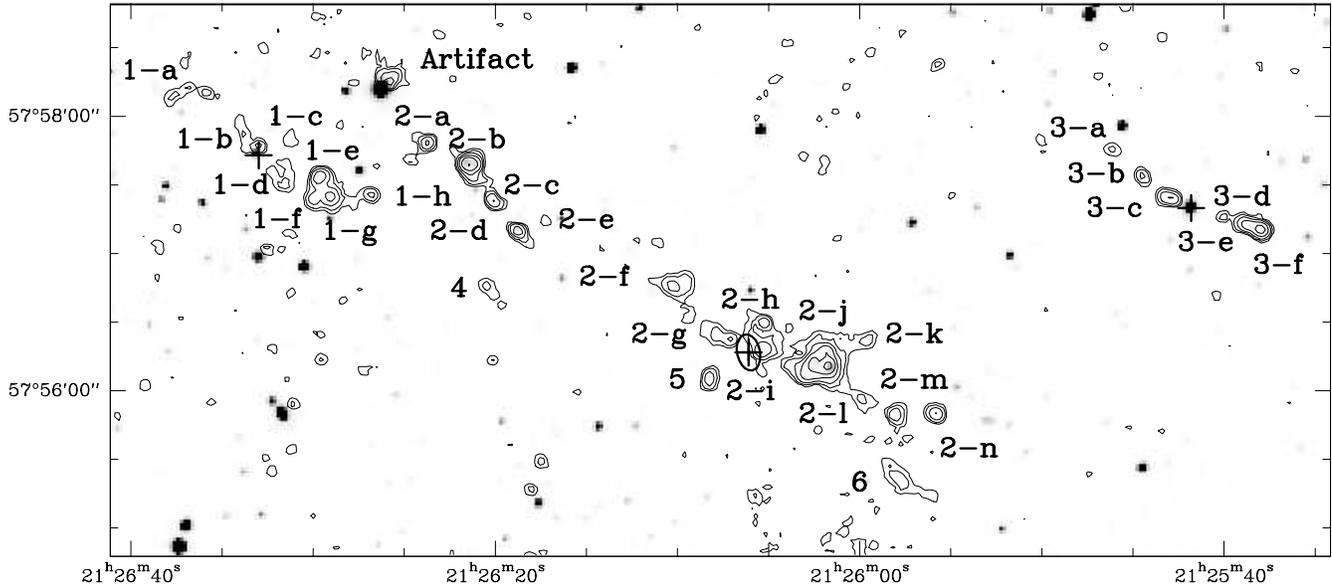}}
\caption{Blow-up of the inner region of the observed field. In grayscale the
H$_2$ 1\,--\,0\,S(1) + continuum emission is shown. For clarification the
continuum subtracted H$_2$ image is overlaid in contours for the identification
of the H$_2$ emission. The driving sources are marked by a cross. Contour
levels start at 4\,$\times$\,10$^{-20}$\,W\,m$^{-2}$\,arcsec$^{-2}$ and
increase by a factor of two.}
\label{flows_ic1396w}
\end{figure*}

In Fig.\,\ref{field_ic1396w} we marked the positions of the red sources listed
in Table\,\ref{redstars} by crosses. The majority of these objects is situated
in the south-east corner of the globule (marked by a circle). However, no signs
of outflow activity are visible here. Exactly in this region, Froebrich et al.
(\cite{fshe03}) discovered with ISOPHOT maps an extremely cold source, possibly
cold dust. This dust might cause the red colours of the stars and the
non-detection of H$_2$ emission in this field. Still the possibility of other
deeply embedded sources exists.

About 120\arcsec\, south of outflow number 1, a bright nebulosity is visible.
Westward of this object there is also a small group (ellipse) of red sources
(including the reddest of our objects, \#31 in Table\,\ref{redstars}). No signs
of outflow activity is visible here also. The other red stars are distributed
uniformely in the field of the globule.

In our continuum subtracted NB2122 images of the IC\,1396\,W globule we
discovered a variety of emission line knots in the 1\,--\,0\,S(1) line of
molecular hydrogen. In Fig.\,\ref{field_ic1396w} the whole observed field is
shown in the NB2122 image. We marked and labelled the three detected outflows
and regions with stars of extreme red colours. Fig.\,\ref{flows_ic1396w} shows
a blow up of the inner part of our image. The continuum subtracted H$_2$ image
is overplotted in contours for clarification. Contour levels start at
4\,$\times$\,10$^{-20}$\,W\,m$^{-2}$\,arcsec$^{-2}$ and increase by a factor of
two in each step. The detected H$_2$ knots are labelled. Their positions and
fluxes are listed in Table\,\ref{positions}.

The positions of the emission knots strongly suggest that we see three
different outflows, emanating from young stars in the IC\,1396\,W globule. The
knots are named, as shown in Fig.\,\ref{field_ic1396w}, by the outflow (1, 2,
or 3) and a knot number (a, b, c\,...). Three further knots (4, 5, and 6) could
not be associated directly with one of the flows. Surprisingly all three
outflows seem to be orientated from north-east to south-west. The individual
position angles are 58, 60 and 61$^\circ$ for the three individual outflows.
Such a close alignment of three flows (even if just in projection) is difficult
to explain by coincidence (probability: 0.03\,\%). There are also other
examples of aligned outflows in small ($<$\,1\,pc) star forming clouds (e.g.
\object{$\rho$\,Oph\,A}, Kamazaki et al. (\cite{kshuk03}); \object{L\,1551},
Saito et al. (\cite{skimhhk_etal95}); IC\,1396\,N, Nisini et al.
(\cite{nmvgldc_etal01})). The authors state magnetic fields, density gradients
or initial angular momentum of the molecular cloud as cause for the alignment.
A consistent interpretion based on magnetic fields, however, is not possible
for all these objects, since cases exist where fields and outflows are either
parallel (L\,1551, Saito et al. \cite{skimhhk_etal95}) or perpendicular
($\rho$\,Oph\,A, Kamazaki et al. \cite{kshuk03}). Thus, a determination of the
real cause for the aligned outflows is yet not possible with the present data.

\subsection*{Outflow 1}

In a small extension at the north-eastern edge of the IC\,1396\,W globule a
small flow with a projected length of about 90\arcsec\, (0.3\,pc) is detected.
Along the outflow axis a red star (\#9 in Table\,\ref{redstars}) is situated
which possibly drives the flow. According to the K-band magnitude of 12.4\,mag
the object seems to be a late type Class\,1 or a borderline Class\,1/2 source
(Aspin et al. \cite{asr94}). Southwest of this source a group of four (1-d ..
1-g) emission knots is observed. Their positions suggest that they might be
part of a bow shock heading to south-west. About 20\arcsec\, westward of this
emission, another faint knot (1-h) is found. The north-eastern part of the flow
is much fainter in the H$_2$ emission. A careful inspection of the H$_2$ image
reveals that the emission indicates a cavity, created by a jet. A similar
structure can be seen in other young outflows (e.g. \object{HH\,211};
Mc\,Caughrean et al. \cite{mrz94}).

\subsection*{Outflow 2}

The most luminous and largest
(240\arcsec\,{\small$^\wedge\!\!\!\!_=$}\,0.9\,pc) outflow is situated right in
the centre of the observed field. At the south-west end, a bow shock structure
can be found. This part of the emission was already detected in the K' image of
Hodapp (\cite{h94}). Two further knots are visible in the prolongation (2-m,
2-n). In the north-east the flow shows both a cavity-like structure (2-f .. h)
and further knots (2-a .. e), seeming to be part of a bow shock. Between the
knots 2-g .. i, and right at the axis of the flow, the source IRAS21246+5743 is
situated. No other red object can be found along the flow axis. If
IRAS21246+5743 is the source of the flow, the south-west part is much less
extended. This could be due to higher extinction in this direction, since more
red stars and sub-mm emission (Froebrich et al. \cite{fshe03}) can be found in
the prolongation of the flow (see Sec.\,\ref{photometry}). At least a 30\,mag
higher visual extinction would be necessary to avoid the detection of H$_2$
knots of the same brightness as the north-east lobe of the flow. A more
favourable possibility is that the flow enters at its south-western edge a more
dense region. The resulting shock waves are stronger, causing the bright knot
2-j and the flow is prevented to extend more into this direction.

\subsection*{Outflow 3}

The third outflow is the faintest detected in the field. It is situated at the
north-western edge of the globule. The emission knots show a jetlike structure,
consisting of three knots on each side. The jet and counter-jet are misaligned
by a projected angle of 25\,$^\circ$. While the south-western part is more
luminous, it is less extended (SW:
36\arcsec\,{\small$^\wedge\!\!\!\!_=$}\,0.13\,pc; NE:
40\arcsec\,{\small$^\wedge\!\!\!\!_=$}\,0.15\,pc). Similar to flow number 1, a
red star (\#2 in Table\,\ref{redstars}) possibly driving the jet, is found in
the centre of the emission knots. The measured K-band brightness of 10.8\,mag
classifies this source as a Class\,2 object (Aspin et al. \cite{asr94}). This
would also explain the low luminosity of the outflow (see below).

\subsection*{Other H$_2$ emission and flow energetics}

\label{energy}

Three more emission knots are detected in the field that cannot be directly
associated with one of the flows. Two of them (5 and 6) could possibly belong
to the flow driven by IRAS21246+5743. Another possibility is the existence of a
fourth outflow, also parallel to the others, consisting of these three knots. A
possible driving source of the flow, confirming this hypothesis, is not
detected in our NIR images.

The total flux in the 1\,--\,0\,S(1) line of the three outflows is 96, 391 and
71\,$\times$\,10$^{-18}$\,W\,m$^{-2}$. Due to overlap effects in the knots,
this number is not equal to the sum of the fluxes given in
Table\,\ref{positions}. The fluxes convert to 1.7, 6.8 and
1.2\,$\times$\,10$^{-3}$ solar luminosities, using a distance of 750\,pc.
Taking an average extinction of A$_V$\,=\,10\,mag (read off the colour-colour
diagram, Fig.\,\ref{fhd_ic1396w}) and A$_V$/A$_{2.1\mu m}$\,=\,9 this converts
to 5, 19 and 3\,$\times$\,10$^{-3}$\,L$_\odot$ total intrinsic H$_2$
1\,--\,0\,S(1) luminosity. According to the classification of the driving
sources (flow\,1: Class\,1/2; flow 2: Class\,0; flow 3: Class\,2), these
measurements fit well in the evolutionary picture of outflow and source
luminosities (e.g. Smith \cite{s99, s00, s02}). 

IRAS21246+5743, the driving source of outflow 2, is possibly of Class\,0
(Froebrich et al. \cite{fshe03}) and has a luminosity of about 16 solar
luminosities. Froebrich et al. (\cite{fshe03}) further estimate a final source
mass of about one solar mass. In their Fig.\,13 they present evolutionary
tracks for the outflow and source luminosity. This leads to an estimate of the
total shock luminosity of about 14\,L$_\odot$, and thus (taking 10\% of the
emission in H$_2$ lines, 10\% of these in the 1\,--\,0\,S(1) line and
A$_V$\,=\,10\,mag) to 12\,$\times$\,10$^{-3}$\,L$_\odot$ in the 1\,--\,0\,S(1)
line. Here we measure 19\,$\times$\,10$^{-3}$\,L$_\odot$ for the whole outflow.
Taking into account the 'unusual' conditions of the south-western lobe (see
Sec.\,\ref{morphology}, Outflow\,2) and calculating the luminosity by using
twice the north-eastern lobe we get 14\,$\times$\,10$^{-3}$\,L$_\odot$. This is
in very good agreement with the model predictions, giving further strong
evidence of IRAS21246+5743 being a Class\,0 object.


\section{Conclusions}

We have observed the IC\,1396\,W globule in J, H, K', and the NB2122 narrow
band filter centred on the 1\,--\,0\,S(1) line of molecular hydrogen. By means
of the continuum subtracted NB2122 image we were able to detect three molecular
outflows in the field. Two of them were not previously known. The flow axes are
parallel within 3\,degrees in projection. More and more such aligned outflows
are discovered in small ($<$\,1\,pc) star forming regions/globules. Magnetic
fields cannot consistently explain all these cases. Thus, a parallel initial
angular momentum of these objects, caused by the fragmentation of small
clouds/globules, might be the reason for the alignment.

With our NIR photometry, IRAS and ISOPHOT observations we are able to identify
the driving sources of the outflows. Two flows are driven by more evolved
Class\,1/2 objects. The brightest outflow is driven by the Class\,0 source
IRAS21246+5743. We measured an intrinsic H$_2$ luminosity of
19\,$\times$\,10$^{-3}$\,L$_\odot$ which confirms the Class\,0 nature of this
object. The other outflow luminosities fit well into the evolutionary model of
Smith (\cite{s99, s00, s02}) for the evolution of outflow and source
luminosities.

The asymmetry in brightness and lenght of the main outflow, driven by
IRAS21246+5743, can be well explained in consideration of the findings of
Froebrich et al. (\cite{fshe03}). At the south-western end the flow enters a
clump of denser material that causes stronger shocks with brighter H$_2$
emission and at the same time decelerates the jet.

Our JHK photometry of the globule reveals a population of young stars. They are
situated mainly in a dense embedded subcluster, about 2.5\arcmin\, south-west
of IRAS21246+5743. This cluster coincides with a clump of denser gas. The other
young stars are almost uniformely distributed in the observed field. The spread
of the red sources over almost the whole globule indicates that star formation
takes place uniformely distributed. It is concentrated, nevertheless, in a
region with strong FIR emission.


\begin{acknowledgements}

It is a pleasure to thank A.P. Hatzes and J. Eisl\"offel for providing
observing time within the framework of their campaign. We greatfully
acknowledge the support from H. Linz and A. Sasse, as well as valuable comments
from the anonymous referee. A. Scholz received travel fund from the DFG
HA3279/2-1 project.

\end{acknowledgements}

\end{document}